\documentstyle[aps,prl,epsf]{revtex}
\draft
\begin{document} 
\preprint{NSF-ITP-97-122, \today}
 \title{Edge Magnetic Field in the xxz Spin-1/2 Chain} 
\author{Ian Affleck}
\address{Department of Physics and Astronomy and Canadian Institute for Advanced
Research, \\ University of British Columbia, Vancouver, BC, V6T 1Z1,
Canada} \maketitle \begin{abstract} 
The critical behavior associated with a transverse magnetic field applied  at the edge of 
a semi-infinite xxz S=1/2 chain is calculated using field theory techniques.  Contrary to 
a recent claim, we find that the long-time behavior is given by a renormalization group 
fixed point corresponding to an infinite field which polarizes the spin at the edge.  The 
zero temperature entropy and position-dependent magnetization are calculated.
\end{abstract} 
There has recently been considerable interest in boundary critical behavior associated 
with various quantum impurity problems.  In particular, it was shown by Kane and 
Fisher \cite{Kane} that even a weak scattering potential in a repulsive Luttinger liquid 
effectively renormalizates to infinity at low energies so that the transmission coefficient 
vanishes.  Independently, the essentially equivalent problem of a single impurity in an 
S=1/2 Heisenberg chain, which is equivalent to the spinless Luttinger liquid, was 
studied \cite{Eggert}.  It was concluded that an arbitrarily small weakening of a single 
exchange coupling causes this coupling to renormalize to 0 giving  a fixed point 
corresponding to a broken chain.  In the bosonization approach, the infrared fixed point 
corresponds to a Dirichlet boundary condition, $\phi (0)=\hbox{constant}$,  on the 
boson field.  It is generally believed that only in the case of a spinful Luttinger liquid, 
where two boson fields must be introduced, does a non-trivial fixed point occur.  

A problem which is closely related to these ones involves a transverse magnetic field, 
$h$, applied at the end of a semi-infinite chain with a free boundary condition:
\begin{equation}
H = J\sum_{i=0}^\infty [S^x_iS^x_{i+1}+S^y_iS^y_{i+1}+\gamma S^z_iS^z_{i+1}]-
hS^x_0.\label{Ham}
\end{equation}
The integrability of this model was shown in \cite{Mccoy}.
The continuum limit of this model, given below, can be written in terms of a boundary 
sine-Gordon field theory and describes a particle in a periodic potential coupled to a 
dissipative environment \cite{Guinea1}.  In the case $\gamma =0$, the connection of 
the dissipation model with the spin-chain problem was exploited in \cite{Guinea2}, 
leading to a mapping into a free electron problem and thus an  explicit solution in the 
continuum limit.  The integrability of the boundary sine-Gordon model was shown in 
\cite{Ghoshal}.
Further work has appeared on the continuum limit bosonized form 
\cite{Callan,Polchinski}.
The connection of this boundary sine-Gordon model with the problem of an impurity in 
an infinite Luttinger liquid was pointed out in \cite{Kane} and \cite{Wong}.  The 
integrability of the boundary sine-Gordon model was used to obtain exact results on the 
impurity in the infinite Luttinger liquid in \cite{Fendley}.   It was recently claimed 
\cite{Oreg} that both the $h=0$ and $h=\infty$ fixed points in the spin chain model are 
unstable and instead, the system renormalizes to some sort of non-trivial intermediate 
$h$ fixed point analogous to the behavior in the overscreened Kondo problem.  If true, 
this would have important consequences for both the Luttinger liquid and quantum 
dissipation problems.

The purpose of this note is to solve for the critical properties of this spin-chain problem 
by an extension of the methods used in \cite{Eggert}.  We conclude that the infinite 
field fixed point is stable along the entire xxz critical line $-1<\gamma <1$.  This fixed 
point corresponds to a boundary condition $S_0^x=\hbox{constant}$.  In bosonization 
language this corresponds to a Neumann boundary condition, $d\phi /dx(0)=0$.    At the 
isotropic point, $\gamma =1$, the edge field is exactly marginal and a line of fixed 
points occurs.  Our conclusion is consistent with the earlier calculations of Guinea et al. 
\cite{Guinea1,Guinea2} but disagrees with the recent results of \cite{Oreg}. We argue 
that the different conclusion  reached in \cite{Oreg} was due to a misinterpretation of 
the precise meaning of the infinite field fixed point.  As further applications of the 
Neumann boundary condition, we calculate the zero temperature impurity entropy and 
$<S^x_j>$, showing that the latter exhibits universal oscillations which decay into the 
chain with a power law. 

The standard bosonization technique (see for example \cite{Affleck}) allows us to 
represent the spin operators in terms of a boson field, $\phi$ with Lagrangian density:
\begin{equation} {\cal L}=(1/2)[(\partial_t\phi )^2-(\partial_x\phi )^2].
\end{equation}  (We set the spin-wave velocity to 1.)
The long time and distance behavior of the spin operators corresponds to separate 
uniform and staggered components:
\begin{eqnarray}
S^z_j&\approx&(1/2\pi R)\partial \phi /\partial x + A(-1)^j\sin \phi /R \nonumber \\
S^-_j&\approx& e^{i2\pi R\tilde \phi}\left[ B\cos \phi /R+C(-1)^j\right].\label{bos}
\end{eqnarray}
Here $\tilde \phi$ is the dual field, defined by splitting $\phi$ up into left and right 
moving terms:
\begin{equation}
\phi (t,x)=\phi_L(t+x)+\phi_R(t-x),\ \   \tilde \phi (t,x)=\phi_L(t+x)-\phi_R(t-x).
\end{equation}
$\phi$ is regarded as an angular variable on a circle of radius $R$ given in terms of the 
exchange anisotropy parameter, $\gamma$, by:
\begin{equation} R=\sqrt{(1/2\pi )-(1/2\pi^2)\cos^{-1}\gamma }.\end{equation}
Along the xxz critical line:
\begin{equation} 0<R<1/\sqrt{2\pi}.\end{equation}
Only the first amplitude in Eq. (\ref{bos}) is a universal function of $R$.  The other 
constants $A$, $B$, $C$ are non-universal.  With an appropriate choice of ultraviolet 
regularization scheme (and the lattice spacing and spin-wave velocity set to 1),
\begin{equation} <S^x_iS^x_j>\to {C^2(-1)^{i-j}\over 2|i-j|^{2\pi R^2}}.
\label{bulk}\end{equation}
Thus the ($\gamma$-dependent) constant, $C$, can be determined from a numerical 
calculation of this correlation function.  An exact result for the amplitude of the 
correlation function was conjectured recently \cite{Lukyanov}:
\begin{equation}C(\gamma ) = {(1+\xi )\over 2}\left[{\Gamma ({\xi \over 2})\over 
2\sqrt{\pi}\Gamma \left({1\over 2}+{\xi \over 2}\right)}\right]^{\eta /2}\times \exp 
\left\{-(1/2)\int_0^\infty {dt\over t}\left({\sinh (\eta t)\over \sinh (t) \cosh[(1-\eta )t]}-
\eta e^{-2t}\right)\right\},\label{exact}\end{equation}
where $\eta \equiv 2\pi R^2$ and $\xi \equiv \eta /(1-\eta )$.
This function is plotted in Figure 1.  [It goes to .5 at $\gamma \to -1$ and diverges as 
$(1-\gamma)^{-1/8}$ as $\gamma \to 1$.]  As we shall see below, the same constant 
$C$ will apear in  $<S^x_j>$ in the presence of a boundary field.  
\begin{figure}
\epsfxsize=10 cm
\epsfbox[20 250 500 550]{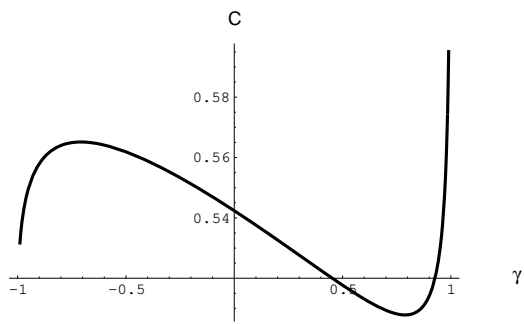}
\caption{The amlplitude, $C$ appearing in the bosonization formula of Eq. 
(\protect{\ref{bos}}), as a function of the anisotropy parameter, $\gamma$, defined in 
Eq. (\protect{\ref{Ham}}).}
\label{fig:C}
\end{figure}

It was shown in \cite{Eggert} that free boundary conditions on the spins at $x=0$ 
correspond to a Dirichlet boundary condition on the boson field, 
\begin{equation}\phi (0)=0.\end{equation}
(The value of the constant follows from requiring that $<S^z_j>=0$.)
To calculate Green's functions or RG flows at this fixed point we regard this boundary 
condition as relating $\phi_L$ and $\phi_R$:
\begin{equation}\phi_R(t,0)=-\phi_L(t,0).\end{equation}
In fact, this boundary condition allows us to regard $\phi_R(x)$ as the analytic 
continuation of $\phi_L(x)$ to the negative $x$ axis:
\begin{equation} \phi_R(t,x)=-\phi_L(t,-x).\end{equation}
This allows a straightforward evaluation of Green's functions \cite{Eggert}.

Now we consider the effect of the edge magnetic field.  The staggered component of 
$S^x_0$ gives an extra boundary term in the free boson Hamiltonian:
\begin{equation}
H_B= -\hbox{constant}\cdot h \cos 2\pi R\tilde \phi (0).\label{bosint}\end{equation}
Using the fact that we have a free boundary condition, we may equivalently write this 
as:
\begin{equation}
H_B= -\hbox{constant}\cdot h \cos [4\pi R \phi_L(0)] .\end{equation}
This has a scaling dimension:
\begin{equation} d=2\pi R^2.\end{equation}
Since $d<1$ along the entire xxz critical line this is a relevant boundary interaction.  At 
the antiferromagnetic Heisenberg point $\gamma =1$, it is marginal.  We return to this 
special case below.  The simplest possibility to assume is that $h$ renormalizes to 
$\infty$ in the infrared.  From Eq. (\ref{bosint}) this implies a Neumann boundary 
condition
\begin{equation} \tilde \phi (0)=\phi_L(0)-\phi_R(0)=0, \end{equation} 
and hence $\partial \phi /\partial x=\partial \tilde \phi /\partial t=0$.  What this 
assumption means in practical terms is that Green's functions involving spatial locations 
far from the chain-end compared to a crossover length scale $\xi$ (given below) and 
long time intervals compared to $1/\xi$ will be given by the free boson model with the 
Neumann boundary condition.  To check the consistency of this assumption, we should 
calculate the scaling dimension of all operators allowed by symmetry which could be 
added to the effective Hamiltonian.  Our assumption is consistent if these are all 
irrelevant, $d>1$.  Imposing the boundary condition, we may write the spin operator at 
the origin as:
\begin{equation} 
S^x_0 \approx C+ B\cdot  \cos (2\phi_L/R)
\end{equation}
The operator $\cos (2\phi_L/R)$ is certainly allowed by symmetry in the effective 
Hamiltonian, since it occurs in $S^x_0$.  It has scaling dimension:
\begin{equation} d=1/2\pi R^2 >1,\end{equation} and is therefore irrelevant. This is a 
natural result given the interpretation of the Neumann boundary condition as 
corresponding to infinite field.  Applying an additional field at any sites near the chain 
edge shouldn't destabilize the fixed point.

The marginal operator $\partial \tilde \phi /\partial x$ is forbidden by the symmetry of 
rotation by $\pi$ around the x-axis, which is still a good symmetry in the presence of 
the magnetic field.  This takes $\tilde \phi \to -\tilde \phi$.  Thus only irrelevant 
operators, $(\partial \tilde \phi /\partial x)^2$, $\cos 2n\phi_L/R$ (for $n=1,2,3, \ldots$) 
are allowed.
Hence the infinite field fixed point is stable.  It is thus very natural to assume that even 
a very small field will renormalize to $\infty$ so that the Neumann boundary condition 
describes the long-time behavior.  Indeed it is difficult to imagine what a non-trivial 
fixed point (neither Dirichlet nor Neumann) would look like.  In fact, it can be proven 
that Dirichlet and Neumann boundary conditions are the only conformally invariant 
ones in a theory containing a single periodic boson, for generic $R$ \cite{Friedan}.  
Hence, there can be no other fixed points if we assume that they correspond to 
conformally invariant boundary conditions.

Note that if we had ignored the Neumann boundary condition and considered the bulk 
operator $\cos (\phi /R)$ we would have obtained the dimension $1/4\pi R^2$, which is 
precisely 1/2 the correct value.  This obeys $d<1$ along the xxz critical line for 
$\gamma >0$ which would imply that the operator was relevant and the infinite field 
fixed point was unstable.  The infinite field limit {\it does not} give a problem 
equivalent to the initial one because the boundary conditions have changed from 
Dirichlet to Neumann.  The result is that a magnetic field (in the x-direction) becomes 
irrelevant with Neumann boundary conditions while it was relevant with Dirichlet 
boundary conditions.  It is instructive to contrast the RG behavior of the present 
problem with a spin chain version of the 2-channel Kondo problem of the type treated 
in \cite{Eggert} where we consider the Heisenberg model ($\gamma =1$) on the 
infinite line with a \  coupling, $J_K$  between sites $0$ and $\pm 1$ which is different 
than the bulk coupling, $J$.  In that case the $J_K=\infty$ fixed point really is 
equivalent to the $J_K=0$ fixed point  because the three strongly coupled spins form an 
effective $S=1/2$ spin at $J_K=\infty$ and the neighboring spins on sites $\pm 2$ obey 
free boundary conditions with no coupling to the effective spin in that limit.  This 
follows because the exchange coupling between sites 1 and 2 (and also between sites -1 
and -2) maps the low energy states of the strongly coupled 3-spin complex into high 
energy states, so, using second order degenerate perturbation theory, the effective 
coupling to sites $\pm 2$ is of order $J^2/J_K\to 0$.  By contrast, in the magnetic field 
case the $h=\infty$ limit effectively eliminates the spin $\vec S_0$ but the effective 
field acting on $\vec S_1$ does not go to $0$, but rather to a finite value, $-J/2$.  Thus 
this situation is not equivalent to $h\to 0$, contrary to the statement in \cite{Oreg}.  
Instead, we must regard the infinite field limit as a different fixed point corresponding 
to a Neumann boundary condition in the field theory.

The ``groundstate degeneracy'' (exponential of the zero temperature entropy) for 
Dirichlet and Neumann fixed points is \cite{Oshikawa}
\begin{equation} g_D=1/\sqrt{2\sqrt{\pi}R},\ \  g_N=\sqrt{\sqrt{\pi}R}.\end{equation}
We see that $g_N<g_D$ for all $R<1/\sqrt{2\pi}$, that is along the entire xxz critical 
line.  Thus our assumption of renormalization from Dirichlet to Neumann fixed points 
is consistent with the g-theorem \cite{Affleck2}.

Next, as another application of the Neumann boundary condition, we calculate the long-
distance behavior of $<S^x_j>$.  Other Green's functions can be calculated the same 
way.  We use our bulk bosonization formulas, Eq. (\ref{bos}) and use the Neumann 
boundary condition to regard $\phi_R$ as the analytic continuation of $\phi_L$:
\begin{equation} \phi_R(x)=\phi_L(-x).\end{equation}  Hence:
\begin{equation}
S^x_j\approx \cos 2\pi R [\phi_L(x)-\phi_L(-x)]\cdot
\left\{B\cos \{[\phi_L(x)+\phi_L(-x)]/R\}+C(-1)^j\right\}
\end{equation} Thus
\begin{equation}
<S^x_j> \to C(-1)^j<e^{i2\pi R\phi_L(x)}e^{-i2\pi R\phi_L(-x)}>
={C(-1)^j\over (2j)^{\pi R^2}}.\label{profile}\end{equation}
This formula is only valid outside a cross over length which can be estimated from the 
renormalization group, in the case of a weak field as:
\begin{equation}
\xi \propto (J/h)^{1-2\pi R^2}.\end{equation} 
The amplitude, $C$ is determined in the bulk theory [for example from the behavior of 
the correlation function in the infinite system, Eq. (\ref{bulk})] and is independent of 
the strength of the edge field, $h$.  This reflects the fact that the system flows to a 
universal fixed point regardless of the size of $h$.

Finally, we consider the isotropic Heisenberg antiferromaget, $\gamma =1$, 
$R=1/\sqrt{2\pi}$. In this case the direction of the applied field is immaterial so we 
choose the $z$-direction.  This case turns out to be very similar to the case of arbitrary 
$\gamma$ between -1 and +1 with the field in the z-direction so we consider that more 
general situation.  The magnetic field gives the term in the bosonized Hamiltonian 
(using the Dirichlet boundary condition):
\begin{equation} H_B=-h\alpha \cdot \partial \phi_L/\partial x (0),\end{equation}
where $\alpha$ is a non-universal constant of O(1). Actually, this form of the prefactor 
is only valid at small $h$.  For larger $h$ it must be replaced by some non-universal 
function of $h$.  
 Adding this to the free boson Hamiltonian leaves a free boson theory which can be 
solved exactly.  This perturbation is exactly marginal, leading to a line of fixed points.  
Using the Dirichlet boundary condition to eliminate the right-movers, the full 
Hamiltonian can be written:
\begin{equation} H=\int_{-\infty}^{\infty}(d\phi_L/d x)^2 -h\alpha d \phi_L/dx 
(0).\end{equation}
This boundary term can be adsorbed into a discontinuity of the field $\phi_L$ at the 
origin:
\begin{equation} \phi_L'\equiv \phi_L -h\alpha \theta (x)/2,\end{equation}
where $\theta$ is the step function.  Applying this shift to the Dirichlet boundary 
condition we can write:
\begin{equation} <S^z_j>\to B(-1)^j<\sin (\phi_L'(x)-\phi_L'(-x)+h\alpha 
/2)/R>.\end{equation}
For small $h$ this gives:
\begin{equation}<S^z_J>\propto {h(-1)^j\over |j|^{1/4\pi R^2}}.\end{equation}
The exponent equals $1/2$ at the isotropic point, $2\pi R^2=1$, in agreement with the 
previous calculation, Eq. (\ref{profile}).  Note that in this case the prefactor varies with 
$h$, corresponding to a line of fixed points. The boundary states at the isotropic point 
have been discussed in \cite{Callan}.  In particular, it was shown that the line of fixed 
points terminates at the Neumann fixed point, corresponding to infinite field.  

I would like to thank Masaki Oshikawa for helpful discussions.  This research was 
begun during a stay at the Institute for Theoretical Physics, Santa Barbara.  It was 
supported in part by NSERC of Canada. 

 \end{document}